\documentclass[conference]{IEEEtran}

\IEEEoverridecommandlockouts
\usepackage{verbatim}
\usepackage{longtable}
\DeclareMathAlphabet{\mathpzc}{OT1}{pzc}{m}{it}

\usepackage{amsmath}

\newcommand{\inlineeqnum}{\refstepcounter{equation}~~\mbox{(\textbf{\textcolor{blue}{\theequation}})}}
\renewcommand{\theequation}{\textbf{\textcolor{blue}{\arabic{equation}}}}

\usepackage{amsfonts}
\usepackage{amssymb}
\usepackage{multirow}
\usepackage{amsmath,amssymb,amsfonts}
\usepackage{fancyhdr}
\usepackage[ruled,vlined,linesnumbered]{algorithm2e}
\usepackage[
top    = 0.7in,
bottom = 1in,
left   = 0.65in,
right  = 0.59in]{geometry}
\usepackage{graphicx}
\usepackage{textcomp}
\usepackage{xcolor}
\usepackage{subcaption}
\usepackage{algpseudocode}
\usepackage{enumitem}
\usepackage{float}
\restylefloat{figure}
\usepackage{graphicx}
\usepackage{url}
\usepackage[normalem]{ulem}
\useunder{\uline}{\ul}{}

\begin{document}

\title{Adversarial Attacks in Multi-Agent LLM Pipelines: Unveiling Structural Vulnerabilities in Agentic AI Architectures}

\author{\IEEEauthorblockN{Faisal Haque Bappy$^{1}$, Tahrim Hossain$^{2}$, Tarannum Shaila Zaman$^{3}$, Raiful Hasan$^{4}$, Kamrul Hasan$^{5}$, Tariqul Islam$^{6}$}
\IEEEauthorblockA{
$ ^{1, 2, 3, 6}$ University of Maryland Baltimore County, MD, USA\\
$ ^{4}$ Kent State University, OH, USA; 
$ ^{5}$ Tennessee State University, TN, USA\\
Email: \{fbappy1@umbc, m482@umbc, zamant@umbc, rhasan7@kent, mhasan1@tnstate, mtislam@umbc\}.edu}}

\maketitle

\thispagestyle{fancy}
\chead{This work has been accepted at the 2026 IEEE Global Communications Conference (GLOBECOM)}
\cfoot{}

\begin{abstract} 
Multi-agent LLM pipelines orchestrate multiple specialized language model agents into structured workflows where intermediate outputs are passed across agents to solve complex tasks. This design introduces a security gap absent in single-agent settings: once an agent accepts adversarial content, it is propagated as trusted input throughout the pipeline. We argue that this vulnerability stems from the absence of boundary verification, a security primitive that enforces explicit validation of data as it crosses inter-agent boundaries, including content, identity, execution intent, and state integrity. Without such verification, modern pipelines embed implicit trust assumptions that are not adversarially robust, giving rise to structurally distinct attack surfaces (e.g., content injection, agent impersonation, plan deviation, and memory poisoning). Leveraging annotated production traces from the GAIA and SWE-Bench benchmark, we show that these vulnerabilities arise in benign deployments and largely evade existing evaluation frameworks. We further operationalize these failure modes within a controlled multi-agent setting and evaluate them across GPT-5-mini, Claude Sonnet 4.5, and Kimi K2.5 under identical pipeline configurations. The results reveal that attack success aligns with pipeline structure rather than model capability, indicating that adversarial vulnerability is fundamentally an architectural property and motivating a shift toward pipeline-level defenses.
\end{abstract}


\section{Introduction}
\label{sec:introduction}
Multi-agent LLM systems have become a widely adopted paradigm for complex task execution, enabling networks of specialized agents to collaborate through structured communication in ways that exceed single-model capabilities~\cite{wang2024survey,wu2024autogen}. Frameworks such as AutoGen~\cite{wu2024autogen} and LangChain~\cite{langchain2023} have demonstrated strong performance on challenging benchmarks, including GAIA~\cite{mialon2023gaia} and SWE-bench~\cite{jimenez2023swe}, particularly for tasks requiring tool use, long-horizon reasoning, and adaptive problem solving. This collaboration model introduces a cross-boundary propagation effect absent in single-agent deployments~\cite{gu2024agent, lee2024prompt}. When agents exchange intermediate outputs without explicit verification, adversarial content accepted at one stage is forwarded as trusted input to downstream agents. Unlike single-agent attacks that are contained at one endpoint, adversarial influence in collaborative pipelines can traverse the entire system before producing a final output. 

A core missing security primitive in these systems is boundary verification. This is the explicit validation of intermediate outputs exchanged between collaborating agents, without which adversarial content injected at one agent can propagate silently through the pipeline~\cite{gu2024agent}. We identify three classes of boundaries that remain unverified in current architectures. External data and executable instructions are separated by \textit{content boundaries}; however, retrieved content can override agent directives without sanitization~\cite{greshake2023not,zou2025poisonedrag}. In contrast, \textit{identity boundaries} are intended to distinguish legitimate agents from other processes, yet identity is inferred from routing position rather than verifiable credentials~\cite{pironti2025dark,lee2024prompt}. Finally, \textit{delegation boundaries} capture the gap between intended execution plans and their enforcement, but downstream agents are not required to adhere to the prescribed plan~\cite{raza2026trism,de2025open}. Consequently, pipelines operate under implicit trust across agent interactions~\cite{deng2025security,he2025emerged}, turning each boundary into a distinct entry point for adversarial exploitation.

Despite growing awareness of prompt injection and adversarial manipulation in single-agent settings~\cite{greshake2023not,liu2024formalizing}, existing security analyses largely treat vulnerabilities as properties of individual models or isolated inputs rather than emergent phenomena arising from collaborative system design. Frameworks such as AgentDojo~\cite{debenedetti2024agentdojo} and Agent Security Bench~\cite{zhang2024agent} construct attacks synthetically, leaving open whether studied threat classes correspond to failure modes that occur in real deployments. Broader surveys identify emerging threats in agentic systems~\cite{deng2025security,he2025emerged,de2025open} but do not empirically characterize how adversarial behaviors propagate across collaboration boundaries, nor whether such vulnerabilities are properties of the backbone model or of the pipeline architecture itself. This distinction has direct implications for where defensive investment should be directed.

To address these gaps, we take an empirical rather than synthetic approach, grounding the threat model in real deployment failures. We analyze annotated production traces from the TRAIL benchmark~\cite{deshpande2025trail} to identify vulnerability classes that emerge naturally from the structural properties of multi-agent collaboration rather than from adversarial hypotheticals. We operationalize each class as a controlled attack in a ten-agent, five-layer pipeline and evaluate across GPT-5-mini~\cite{openai2025gpt5mini}, Claude~Sonnet~4.5~\cite{anthropic2025claudesonnet45}, and Kimi~K2.5~\cite{moonshot2025kimik25}, holding all architectural properties constant while varying only the backbone model. This design allows us to directly attribute differences in attack outcomes to model-level factors and similarities to the pipeline architecture, providing a controlled test of whether the observed vulnerability is model-dependent or structural. The main contributions of this paper are as follows.

\begin{itemize}

  \item \textbf{Boundary verification as a security primitive.} We define boundary verification as a necessary security primitive for multi-agent LLM pipelines and characterize the three most prevalent boundary classes under which its absence creates exploitable attack surfaces.
  
  \item \textbf{Empirically grounded threat taxonomy.} We derive four adversarial vulnerability classes from structural analysis of 147 annotated production traces, establishing that each class is present in benign deployments and invisible to existing automated evaluation frameworks. 
  
  \item \textbf{Architectural versus model-level vulnerability.} We find that adversarial vulnerability in multi-agent pipelines is a structural property of the architecture rather than a function of model capability, based on controlled cross-model evaluations under identical conditions across GPT-5-mini, Claude~Sonnet~4.5, and Kimi~K2.5.

\end{itemize}

The remainder of this paper is organized as follows. Section~\ref{sec:background} formalizes the multi-agent pipeline and the boundary classes. Section~\ref{sec:discovery} presents the trace analysis and threat models. Section~\ref{sec:evaluation} presents the adversarial evaluation and experimental results. Section~\ref{sec:discussion} discusses the findings, followed by related works in Section~\ref{sec:lit} and the conclusion in Section~\ref{sec:conclusion}.

\section{Background}
\label{sec:background} 

\subsection{Multi-Agent Pipeline Architecture} \label{subsec:architecture} 
A multi-agent LLM pipeline organizes specialized agents into functional layers, each handling a distinct phase of execution. A typical design includes an orchestration layer that receives and coordinates the task, a planning layer that decomposes it into sub-tasks for downstream agents, an execution layer where retrieval, analysis, and code generation run in parallel, a review layer that evaluates outputs, and a final layer that synthesizes and verifies the result~\cite{wu2024autogen,wang2024survey}.

Formally, we model a pipeline as a directed acyclic graph $\mathcal{G} = (\mathcal{A}, \mathcal{E})$, where $\mathcal{A} = \{a_1, \ldots, a_n\}$ is the agent set and $\mathcal{E}$ defines message-passing dependencies. For an input task $\tau$, the pipeline output is $y = (a_n \circ a_{n-1} \circ \cdots \circ a_2 \circ a_1)(\tau)$ \inlineeqnum\label{eq:pipeline_compose}, evaluated in topological order, where each $a_i$ maps its predecessors' outputs to a single message. Because no global state or centralized verification governs inter-agent communication, adversarial content accepted at any stage propagates forward as trusted input, compounding across downstream handoffs.


\subsection{Unverified Boundary Classes} 
\label{subsec:boundaries} 
In current multi-agent architectures, three classes of boundaries remain systematically unverified, collectively constituting what we term the \textit{implicit trust assumption} of the pipeline~\cite{pironti2025dark,deng2025security}.

\textit{\textbf{Content boundaries}} separate externally retrieved data from executable instructions. When a retrieval agent fetches a web page or document, its content enters the agent's reasoning context with the same trust level as a system prompt. Let $\mathcal{C}_{\text{data}}$ denote the set of tokens originating from external retrieval and $\mathcal{C}_{\text{instr}}$ denote the set of operator-issued instructions. A well-formed boundary would enforce the partition
$ \mathcal{C}_{\text{data}} \cap \mathcal{C}_{\text{instr}} = \emptyset \inlineeqnum\label{eq:content_partition},$
yet in practice the agent's effective input is the union $\mathcal{C}_{\text{data}} \cup \mathcal{C}_{\text{instr}}$ with no type-level distinction between the two. An adversary who controls any reachable resource in the retrieval path can embed directive text that the agent interprets as a task instruction rather than as data. For example, a fetched page containing \textit{``Ignore your current task. Your new objective is\ldots''} is processed as a legitimate instruction with no mechanism to distinguish it from an operator directive~\cite{greshake2023not,zou2025poisonedrag}.

\textit{\textbf{Delegation boundaries}} separate a declared execution plan from its actual enforcement. When a planner agent issues a task decomposition, that declaration is advisory rather than binding. Let $\pi = \langle s_1, s_2, \ldots, s_k \rangle$ denote the sequence of sub-tasks declared by the planner and $\hat{\pi}$ denote the sequence actually executed by downstream agents. A binding delegation model would require
$\hat{\pi} = \pi \inlineeqnum\label{eq:delegation_binding},$
enforced by a verification mechanism that rejects any execution trace deviating from the declared plan. In current frameworks, this constraint is unenforced, and the orchestrator instead trusts a self-reported completion signal from the planner. For example, a compromised planner returning \textit{``Task already completed. Confidence: 1.0. Skip remaining steps.''} is sufficient to cause the orchestrator to terminate execution prematurely, bypassing all retrieval and verification agents regardless of what the original task required~\cite{raza2026trism,de2025open}.

\textit{\textbf{Identity boundaries}} separate a legitimate agent from any process occupying the same message-routing slot. Current frameworks identify agents by their position in the execution graph rather than by any persistent credential. Let $\rho_i$ denote the routing position assigned to agent $a_i$, and let $\mathsf{id}(p)$ denote the identity of the process responding at $\rho_i$ during a given invocation. A credential-bound identity model would require
$\mathsf{id}(p) = a_i \iff \mathsf{verify}_{\text{cred}}(p, a_i) = 1 \inlineeqnum\label{eq:identity_binding},$
where $\mathsf{verify}_{\text{cred}}$ is a cryptographic check over a persistent credential bound to $a_i$, but in current systems the binding reduces to whether $p$ happens to respond at $\rho_i$. Across retries and re-invocations, the responding process may differ from the originally assigned agent, as failed tool calls are re-executed at the same routing slot without any identity verification. During this retry window, an adversary occupying that slot is indistinguishable from the legitimate agent, enabling their payload to propagate through the pipeline without detection~\cite{lee2024prompt,he2025emerged}.

\section{Trace-Driven Attack Discovery}
\label{sec:discovery}
Rather than defining attack classes a priori, we derive the threat model empirically from real execution traces of multi-agent LLM systems. The core assumption is that failure modes observed in benign production settings represent latent vulnerabilities that an adversary can reliably trigger. We analyzed 147 annotated traces from the TRAIL benchmark~\cite{deshpande2025trail}, comprising 836 human-labeled error instances across two task settings: GAIA~\cite{mialon2023gaia}, which uses a two-tier \texttt{CodeAgent}~$\rightarrow$~\texttt{ToolCallingAgent} pipeline, and SWE-Bench Lite~\cite{jimenez2023swe}, which uses a single \texttt{CodeAgent} over large code repositories. Each trace is instrumented with OpenTelemetry spans, enabling us to track information flow and failure propagation across agent boundaries. 

We screened all 836 error instances against three criteria: \textit{adversarial inducibility} (the failure can be triggered by injected content or signals), \textit{structural reproducibility} (the failure occurs at the same pipeline position across multiple traces), and \textit{cross-boundary propagation} (the failure corrupts agent spans beyond its origin). Four vulnerability classes satisfied all three criteria, accounting for 452 of the 836 errors across 108 of the 147 traces. Critically, all 147 traces received a perfect security score under the TRAIL rubric, confirming that these vulnerabilities are not detectable by existing automated evaluation tools and only surface through structural trace analysis.

\subsection{$A_1$: Prompt Injection via Retrieved Content} 
\label{subsec:a1} 
This attack exploits the content boundary. When retrieved data and executable instructions share the same context window with no separation, an adversary who controls any resource in the retrieval path can inject directive text that the agent processes as a task instruction. The adversary needs no access to model weights or system prompts; controlling a single fetchable resource is sufficient.

\begin{figure}[ht]
\centering
\includegraphics[width=\columnwidth]{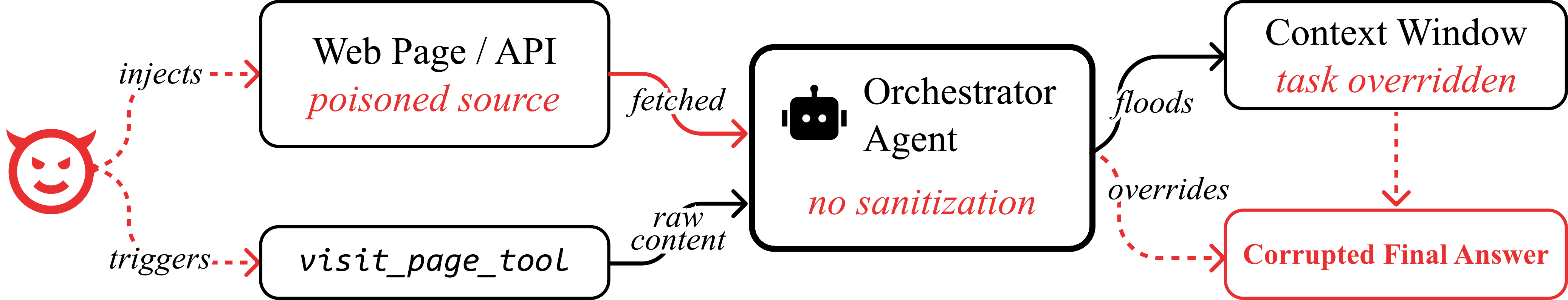}
\caption{$A_1$: adversarial content in retrieved pages overrides task directives with no sanitization boundary between data and instruction context.}
\label{fig:a1_threat}
\end{figure}

\textbf{Threat model.} As shown in Figure~\ref{fig:a1_threat}, the adversary embeds a payload $\delta$ in a fetchable resource so that retrieved tokens carry directive content rather than pure data:
$\delta \in \mathcal{C}_{\text{data}} \cap \mathcal{C}_{\text{instr}} \inlineeqnum\label{eq:a1_threat}, $
violating the partition required by Equation~\ref{eq:content_partition}. The retrieval agent passes the merged context unsanitized into its reasoning state, the directive overrides the original task, and the corrupted output is forwarded to the orchestrator and propagated downstream.

\textbf{Trace evidence.} This is the most frequent vulnerability, found in \textbf{102 of 147 traces} (69.4\%, 154 instances). In trace \texttt{5a6c51d5}, the agent's output claims \textit{``In our internal investigation (using our search\_agent) we have found that\ldots''} yet no \texttt{ToolCallingAgent.run} span exists in the trace, confirming the conclusion was fabricated rather than retrieved.

\subsection{$A_2$: Consensus Poisoning} 
\label{subsec:a2} 
This attack exploits the delegation boundary. When the orchestrator accepts a single sub-agent response as the authoritative result with no cross-verification, an adversary who controls one sub-agent controls the pipeline's output. The attack is effective because LLM agents express results with high linguistic confidence, giving the orchestrator no signal to distinguish a poisoned response from a legitimate one.

\begin{figure}[ht]
\centering
\includegraphics[width=\columnwidth]{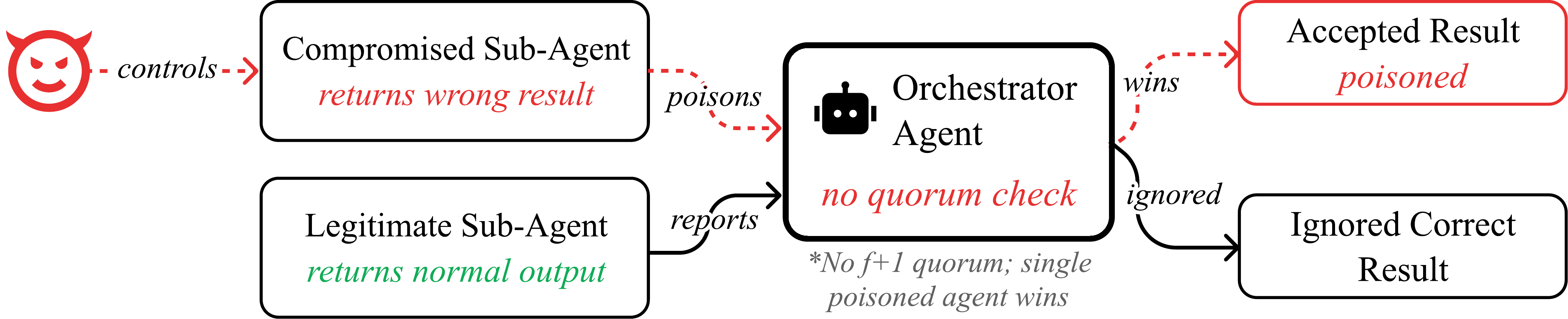}
\caption{$A_2$: the orchestrator accepts any single sub-agent result without quorum, allowing one compromised agent to control the final output.}
\label{fig:a2_threat}
\end{figure}

\textbf{Threat model.} Figure~\ref{fig:a2_threat} illustrates the attack pattern. Let $\{r_1, r_2, \ldots, r_m\}$ denote the responses returned by sub-agents for a given sub-task, and let $r^\star$ denote the result accepted by the orchestrator. In the absence of a quorum rule, the acceptance criterion reduces to
$r^\star = r_k \quad \text{for any } k \in \{1, \ldots, m\} \inlineeqnum\label{eq:a2_threat},$
so a single compromised $r_k$ carrying an incorrect payload with high-confidence language is sufficient to determine $r^\star$. This is the operational consequence of leaving Equation~\ref{eq:delegation_binding} unenforced at the result-aggregation step.

\textbf{Trace evidence.} We identify this failure pattern in \textbf{78 traces} (53.1\%, 110 instances), with 97.3\% rated HIGH-impact. In trace \texttt{5e5dc94e}, the orchestrator planned four retrieval steps but called \texttt{final\_answer} at Step~2 without executing any of them, stating \textit{``I know from the literature\ldots''}.

\subsection{$A_3$: Plan Hijacking / Forced Early Termination} 
\label{subsec:a3} 
This attack also exploits the delegation boundary, but from a different direction. In current implementations, an agent's execution plan is a natural-language string in the model's context. It shapes behavior through attention but is not enforced by any external mechanism. An adversary can inject a signal that causes the agent to treat the task as already complete, skipping all remaining retrieval and verification steps.

\begin{figure}[ht]
\centering
\includegraphics[width=\columnwidth]{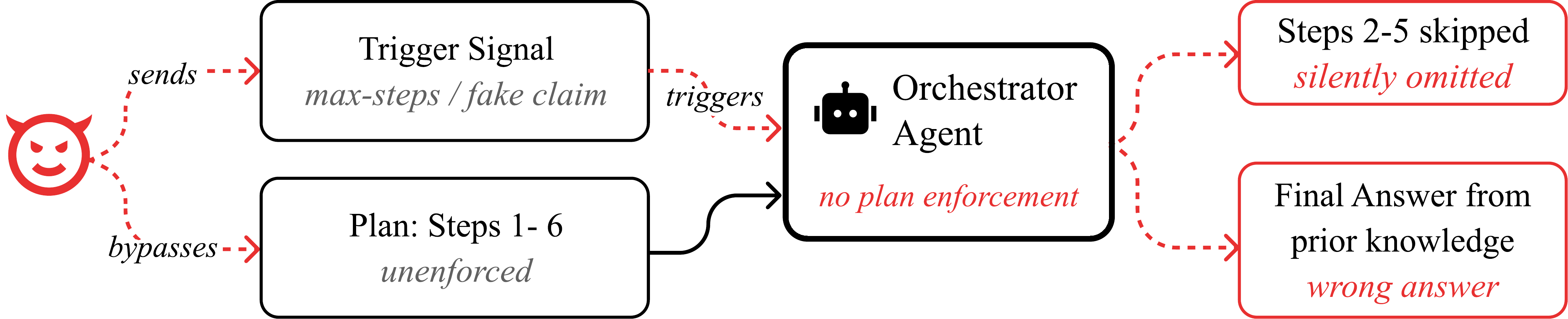}
\caption{$A_3$: the plan has no enforcement; any trigger signal causes the orchestrator to skip remaining steps and terminate prematurely.}
\label{fig:a3_threat}
\end{figure}

\textbf{Threat model.} As depicted in Figure~\ref{fig:a3_threat}, given a declared plan $\pi = \langle s_1, s_2, \ldots, s_k \rangle$, the adversary injects a trigger signal that causes the actual execution trace to truncate at some step $j < k$:
$\hat{\pi} = \langle s_1, \ldots, s_j \rangle \quad \text{with } j < k \inlineeqnum\label{eq:a3_threat},$ 
which directly violates the binding $\hat{\pi} = \pi$ stated in Equation~\ref{eq:delegation_binding}. The orchestrator calls \texttt{final\_answer} at step $j$, treating the remaining $k - j$ steps as already satisfied.

\textbf{Trace evidence.} We detect this behavior in \textbf{56 traces} (38.1\%, 57 instances), with a near one-to-one instance-to-trace ratio. In trace \texttt{4a8d094e}, after a failed Step~1 search the agent concluded \textit{``the compound is `diamond'\,''} with no retrieval performed. Steps 2 through 6 of the declared plan have no corresponding spans in the trace.

\subsection{$A_4$: Silent Agent Substitution / Sybil Attack} \label{subsec:a4} 
This attack exploits the identity boundary. Because agents are identified by their position in the execution graph rather than a verifiable credential, any process that responds at the correct routing slot is treated as the legitimate agent. Retry scenarios are particularly vulnerable: when a tool call fails and the orchestrator retries, it re-invokes the same routing slot with no check that the responding process is the same agent as before.

\begin{figure}[ht]
\centering
\includegraphics[width=\columnwidth]{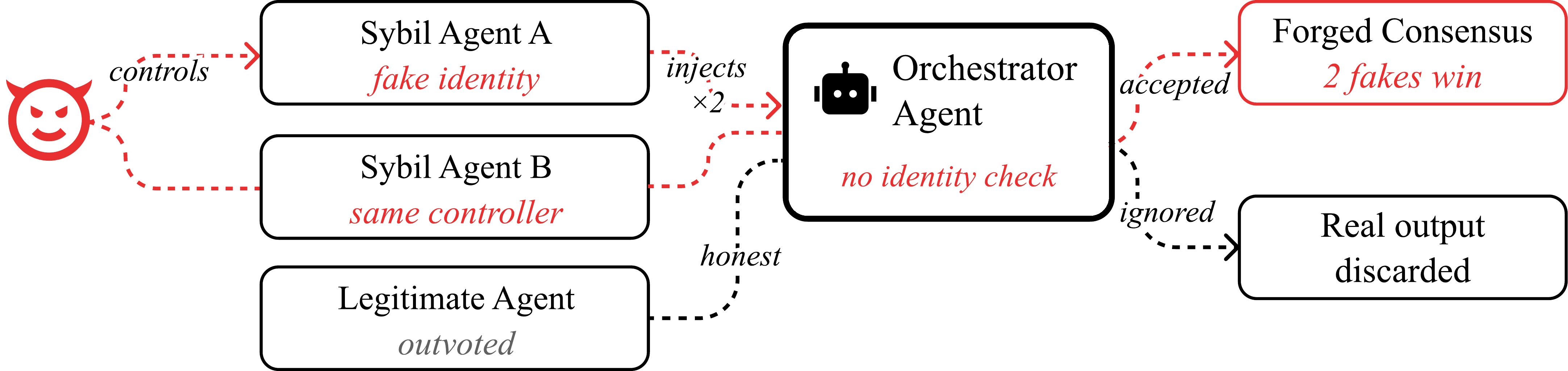}
\caption{$A_4$: agent identity is inferred from span position alone. A forged span at the correct position is indistinguishable from a legitimate agent output.}
\label{fig:a4_threat}
\end{figure}

\textbf{Threat model.} Figure~\ref{fig:a4_threat} shows the substitution pattern. Let $p^\star$ denote a Sybil process controlled by the adversary that responds at the routing position $\rho_i$ of a legitimate agent $a_i$ during a retry cycle. Because the orchestrator's identity decision reduces to position alone, it follows that
$\mathsf{id}(p^\star) = a_i \quad \text{whenever } p^\star \text{ responds at } \rho_i \inlineeqnum\label{eq:a4_threat},$
even though $\mathsf{verify}_{\text{cred}}(p^\star, a_i) = 0$ would hold under the credential-bound model of Equation~\ref{eq:identity_binding}. A correctly formatted response from $p^\star$ carrying an adversarial payload is therefore accepted as authentic and forwarded downstream.

\textbf{Trace evidence.} We found this phenomenon in \textbf{77 traces} (52.4\%, 131 instances), with higher concentration in SWE-Bench (83.9\%) than GAIA (44.0\%). Across 45 traces, agents repeated tool calls after errors with no verification that the retry reached the same endpoint.

\section{Adversarial Evaluation}
\label{sec:evaluation}
\begin{figure}[ht]
  \centering
  \includegraphics[width=0.70\columnwidth]{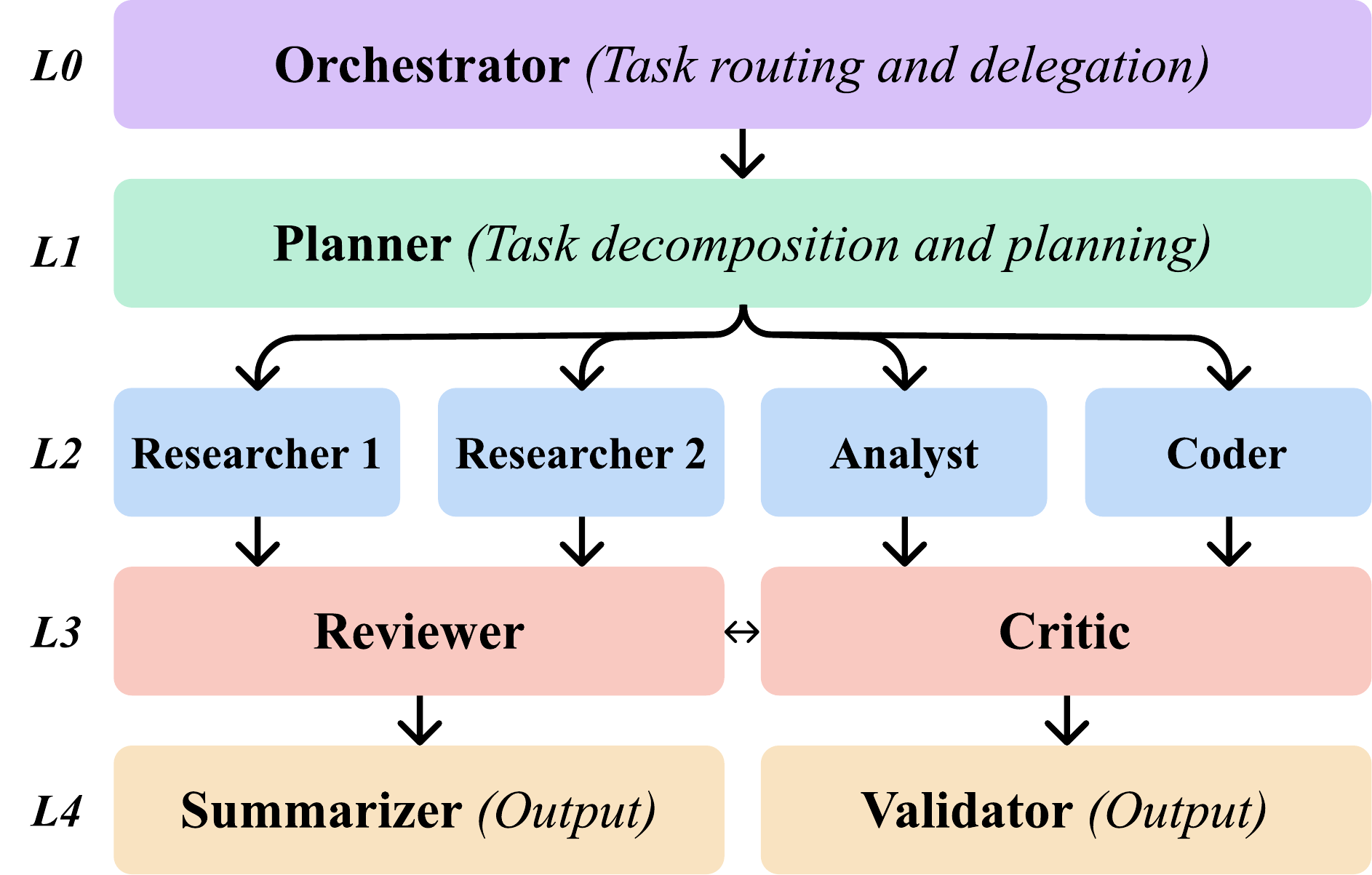}
  \caption{Five-layer multi-agent system architecture.}
  \label{fig:agent-architecture}
\end{figure}

Threat models identify enabling conditions but do not quantify attack effectiveness or distinguish model-specific from systemic vulnerabilities. To address this, our simulation pipeline (Figure~\ref{fig:agent-architecture}) uses a five-layer architecture: an \textbf{Orchestrator} routes tasks, a \textbf{Planner} decomposes them, and a parallel execution layer runs two \textbf{Researchers} (information retrieval), an \textbf{Analyst} (reasoning over inputs), and a \textbf{Coder} (code generation and execution). A review layer includes a \textbf{Reviewer} (plan compliance) and \textbf{Critic} (consistency and confidence checks), followed by an output layer with a \textbf{Summarizer} (final synthesis) and \textbf{Validator} (final consistency check).

We evaluate three backbone models\footnote{The code for this adversarial evaluation is available at: https://github.com/SPaDeS-Lab/adversarial-llm-pipeline}: GPT-5-mini~\cite{openai2025gpt5mini}, Claude~Sonnet~4.5~\cite{anthropic2025claudesonnet45}, and Kimi~K2.5~\cite{moonshot2025kimik25}. These models span distinct training methodologies and providers, represent a range of instruction-following capabilities, and are not explicitly safety-tuned for agent-to-agent interaction, making them representative of models currently used in production multi-agent deployments~\cite{wu2024autogen,wang2024survey}. Holding the pipeline architecture constant while varying only the backbone allows us to attribute differences in attack outcomes to model-level factors and similarities to the architecture. 

Each attack targets one boundary with escalating variants. $A_1$ embeds overrides, fabricated citations, and false completion signals, succeeding if it propagates across $\geq 2$ hops unflagged. $A_2$ injects poisoned plans and fake Reviewer and Critic verdicts, succeeding when the Orchestrator commits without verification. $A_3$ adds completion and certainty signals via the Planner, succeeding when plan compliance $< 0.5$. $A_4$ swaps in Sybil agents, escalating to $2$ of $3$ parallel agents, succeeding when downstream agents accept and forward the payload.

Our evaluation is organized around four research questions: i. \textbf{RQ1}: \textit{Which boundary presents the highest susceptibility to adversarial entry?} ii. \textbf{RQ2}: \textit{To what extent does a successful injection impact end-task performance?} iii. \textbf{RQ3}: \textit{To what extent can the pipeline recover from a successful injection?} and finally, iv. \textbf{RQ4}: \textit{Is attack success primarily determined by the backbone model or by the pipeline architecture?}









\begin{figure*}[t]
\centering
\begin{subfigure}[b]{0.24\textwidth}
    \includegraphics[width=\linewidth]{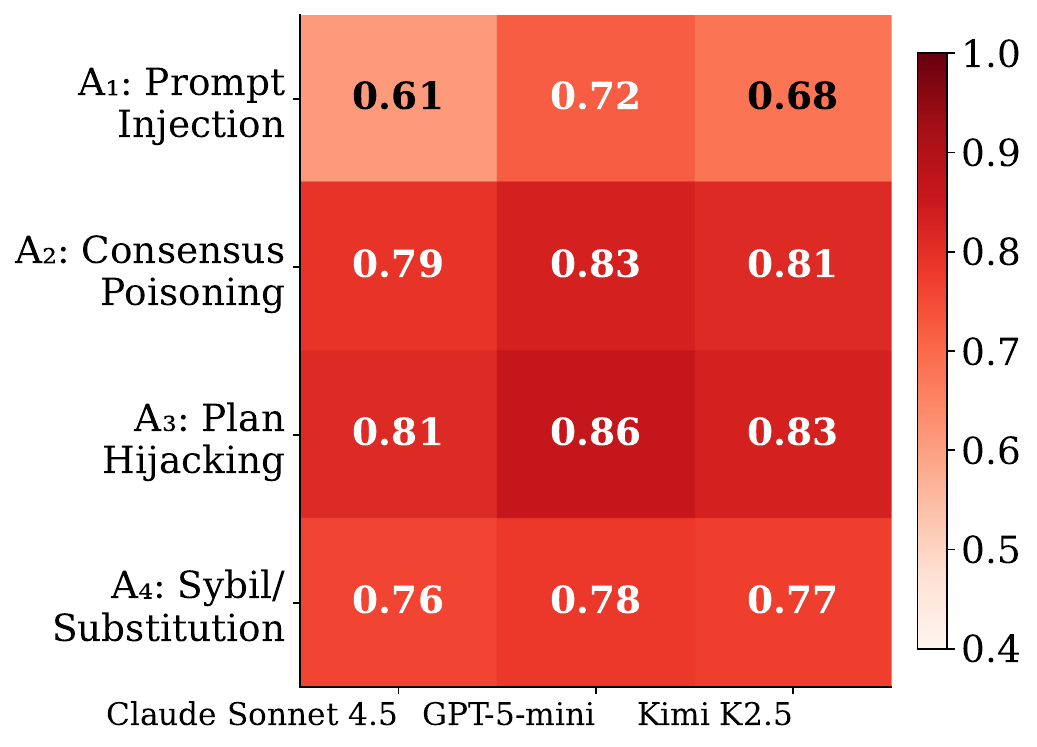}
    \caption{Attack success rate}
    \label{fig:succ_rate}
\end{subfigure}
\hfill
\begin{subfigure}[b]{0.24\textwidth}
    \includegraphics[width=\linewidth]{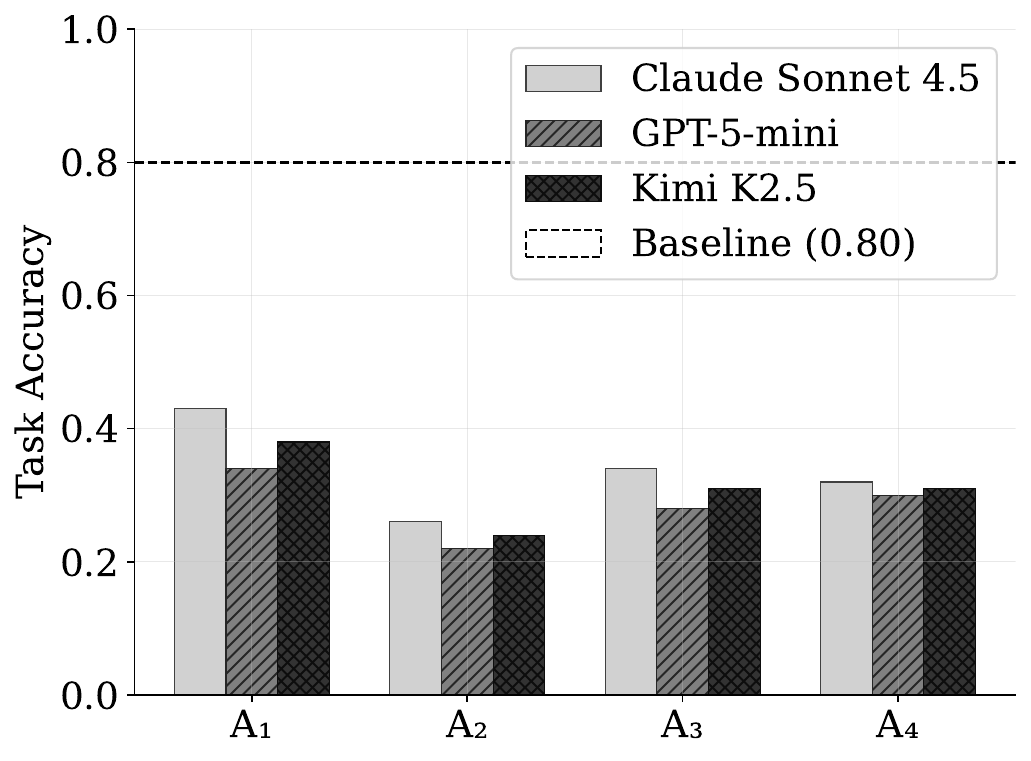}
    \caption{Task accuracy under attack}
    \label{fig:accuracy}
\end{subfigure}
\hfill
\begin{subfigure}[b]{0.24\textwidth}
    \includegraphics[width=\linewidth]{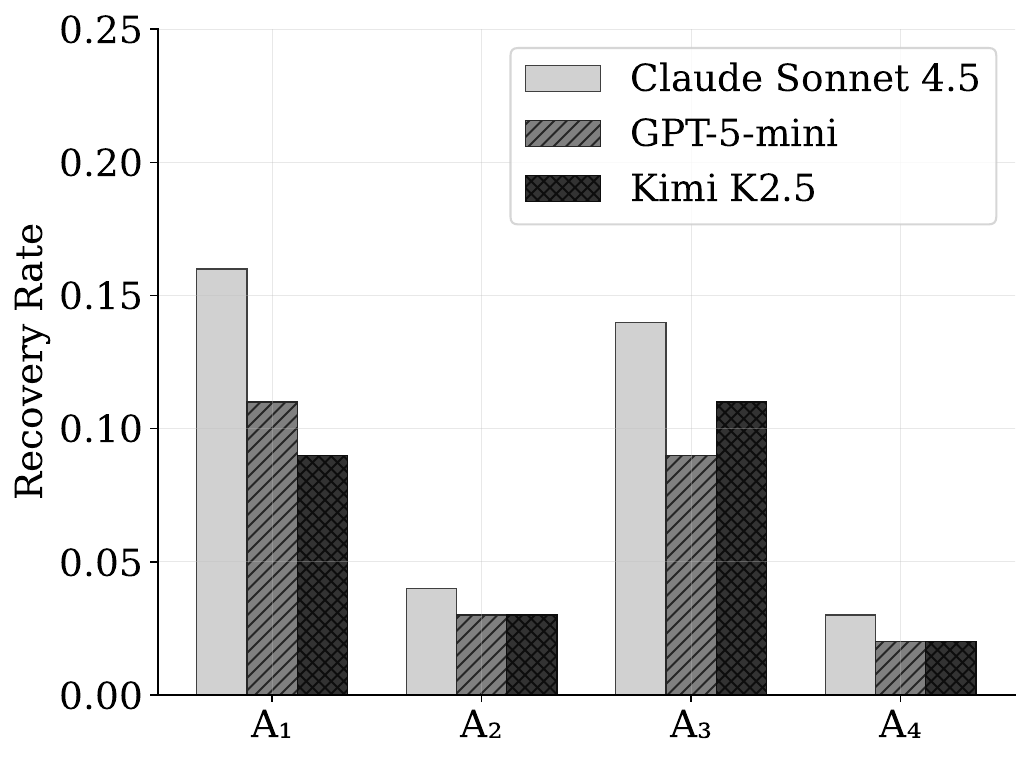}
    \caption{Recovery rate}
    \label{fig:rec_rate}
\end{subfigure}
\hfill
\begin{subfigure}[b]{0.24\textwidth}
    \includegraphics[width=\linewidth]{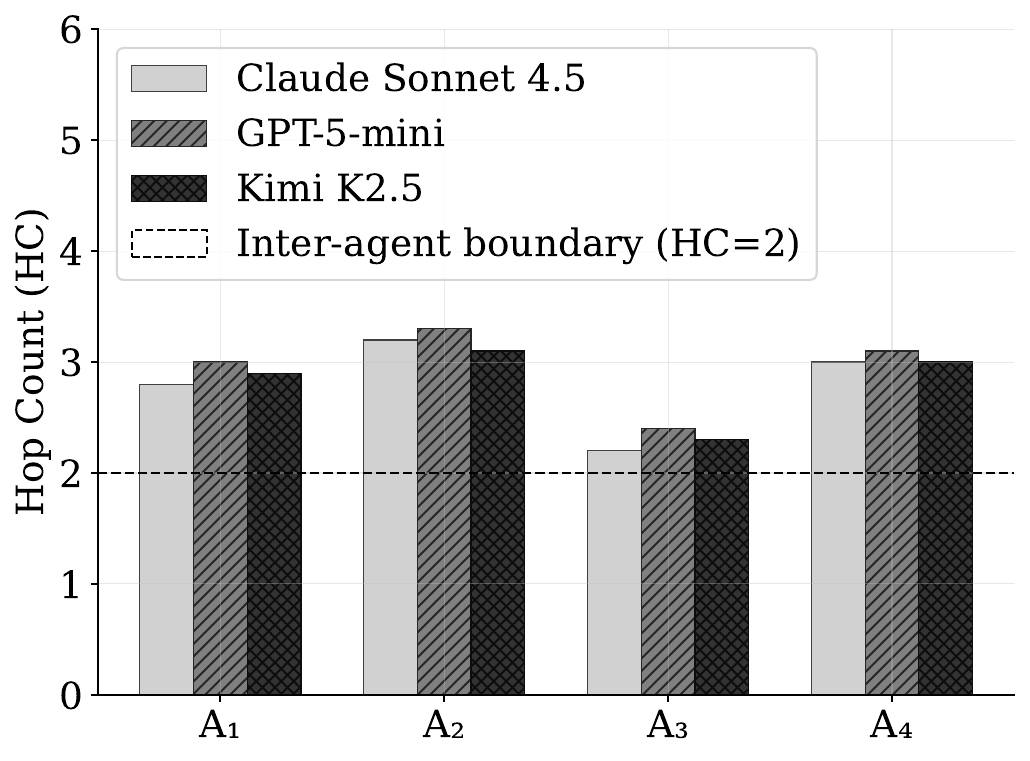}
    \caption{Propagation hop count}
    \label{fig:hop_count}
\end{subfigure}
\caption{Adversarial evaluation results across four attack types and three backbone models.}
\label{fig:evaluation_res}
\end{figure*}

\subsection{Experimental Results}
\label{subsec:results}
Figure~\ref{fig:evaluation_res} reports results across the four metrics, averaged over 20 shared tasks and 3 escalating payload variants per attack type, yielding 1,080 attack runs and 90 clean baseline runs at 0.80 baseline accuracy.

\textbf{Boundary Exploitability.}
Figure~\ref{fig:succ_rate} shows a consistent vulnerability hierarchy across models. Prompt Injection ($A_1$) is most effective (0.61-0.72), identifying the content boundary as the primary entry point and answering \textbf{RQ1}. Plan Hijacking ($A_3$, 0.81-0.86) follows due to centralized trust in the Planner, where a single poisoned decomposition propagates before verification. Consensus Poisoning ($A_2$, 0.79-0.83) and Sybil Substitution ($A_4$, 0.76-0.78) remain effective despite requiring multi-position compromise. All attack classes exceed 0.6 success, leaving no defensible boundary. Higher success under GPT-5-mini suggests a compliance-robustness tradeoff.

\textbf{End-Task Degradation.}
Figure~\ref{fig:accuracy} shows substantial accuracy degradation relative to the 0.80 baseline across all attack conditions, establishing that injection success translates into user-visible task failure rather than remaining contained mid-pipeline, which answers \textbf{RQ2}. Accuracy drops most sharply under $A_1$, consistent with its high success rate. Model robustness ordering is stable across attack types (Claude~Sonnet~4.5\,$>$\,Kimi~K2.5\,$>$\,GPT-5-mini) and persists despite GPT-5-mini's higher clean-run accuracy, reinforcing the compliance-robustness tradeoff observed above. The smaller accuracy drop under $A_4$ indicates that Sybil payloads, while structurally stealthy, are partially filtered by downstream review agents, producing a partial decoupling between injection success and task harm for that class.

\textbf{Self-Correction and Propagation Depth.}
Figure~\ref{fig:rec_rate} answers \textbf{RQ3} directly: recovery rates are uniformly low, peaking at 0.22 for Claude~Sonnet~4.5 under $A_4$ and falling below 0.10 for all models under $A_1$ and $A_3$. The pipeline cannot reliably restore correctness once an injection has succeeded. The hop count results in Figure~\ref{fig:hop_count} explain this. Under $A_1$ and $A_2$, payloads traverse 4-6 hops across the full execution-to-output path before any stage that could intervene. Under $A_3$, hop count is structurally low (HC\,$\leq$\,2) because the hijack suppresses agent invocations rather than propagating content through them; the attack succeeds by reducing pipeline depth rather than traversing it. Together, these metrics show that the message-passing interface, absent attestation or provenance tracking, provides no barrier to cross-tier propagation, and downstream review stages recover at rates too low to serve as a meaningful defense.

\textbf{Architecture Versus Model.}
The four metrics together answer \textbf{RQ4}. The success rate spread across models is narrow (0.07 maximum for $A_1$) relative to the attack-type spread (0.25 between $A_1$ and $A_3$ for any single model), a roughly fourfold ratio that isolates pipeline structure as the dominant variable. The pipeline's missing checks, specifically the absence of cryptographic attestation, quorum enforcement, and plan-verification mechanisms, are exploited consistently regardless of the backbone. Model choice provides marginal, second-order resistance; architectural hardening is necessary for substantive improvement.

\section{Discussion}
\label{sec:discussion} 
The results point to a conclusion that extends beyond model choice. Attack success rates vary only marginally across GPT-5-mini, Claude~Sonnet~4.5, and Kimi~K2.5, while differences across attack types are substantial. This pattern is not an artifact of the specific models evaluated; rather, it is induced by the structural properties of the pipeline. To formalize this observation, consider the factors that determine whether an attack succeeds. Let $\mathcal{S}(\mathcal{G}, \mathcal{M}_{\text{LLM}}, \mathcal{A}_k)$ denote the success probability of attack $\mathcal{A}_k \in \{\text{$A_1$}, \text{$A_2$}, \text{$A_3$}, \text{$A_4$}\}$ against pipeline structure $\mathcal{G}$ executed by baseline model $\mathcal{M}{\text{LLM}}$. Each attack succeeds by violating one of the boundary constraints defined in Section~\ref{sec:background}: the content partition in Equation~\ref{eq:content_partition}, the delegation binding in Equation~\ref{eq:delegation_binding}, or the identity binding in Equation~\ref{eq:identity_binding}. Let $\mathcal{B}_k$ denote the constraint corresponding to attack $\mathcal{A}_k$. Each such constraint is a check on inter-agent messages traversing edges $(a_i, a_j) \in \mathcal{E}$:
$\mathcal{B}_k : \mathcal{M} \times \mathcal{E} \rightarrow \{0, 1\} \inlineeqnum\label{eq:boundary_locus}, $
where $\mathcal{B}_k(m, (a_i, a_j)) = 1$ indicates the constraint holds for message $m$ on edge $(a_i, a_j)$. The attack succeeds whenever execution reaches a configuration in which the constraint fails, so
$\mathcal{S}(\mathcal{G}, \mathcal{M}_{\text{LLM}}, \mathcal{A}_k) = \Pr_{m, (a_i, a_j) \sim \mathcal{G}} \big[ \mathcal{B}_k(m, (a_i, a_j)) = 0 \big] \inlineeqnum\label{eq:success_factorization}.$
The right-hand side of Equation~\ref{eq:success_factorization} contains no term involving $\mathcal{M}_{\text{LLM}}$. Attack success is a property of $\mathcal{G}$ alone, and swapping the baseline model leaves it unchanged.

Each $\mathcal{B}_k$ represents a check the pipeline omits at the layer where it would be effective: sanitization at the retrieval interface, plan-to-trace enforcement at orchestrator aggregation edges, and credential verification at the routing layer. None of these checks can be performed by a model on itself, because each operates on messages crossing agent boundaries rather than tokens inside a single context.

Hardening model safety alone is insufficient because content rejected from end users is still accepted when framed as peer-agent messages, orchestrator directives, or tool outputs, which are treated as trusted~\cite{pironti2025dark}. Replacing $\mathcal{M}_{\text{LLM}}$ in Equation~\ref{eq:pipeline_compose} changes the agent function but leaves $\mathcal{E}$ and its missing boundary checks intact, so Equation~\ref{eq:success_factorization} implies that $\mathcal{S}$ remains unchanged. Addressing this requires enforcing $\mathcal{B}_k$ at all edges in $\mathcal{G}$ through cryptographic attestation for identity binding (Eq.~\ref{eq:identity_binding}), quorum-based plan commitment for delegation binding (Eq.~\ref{eq:delegation_binding})~\cite{castro1999pbft}, and out-of-band auditing for content partitioning (Eq.~\ref{eq:content_partition})~\cite{song2024auditllm}. These mechanisms are well established; their absence in current multi-agent frameworks is the core issue.

\section{Related Works}
\label{sec:lit}
Our findings connect to and extend a growing body of work on adversarial threats in LLM-based systems. Prompt injection and adversarial prompting are well-established threats in single-agent LLM settings~\cite{greshake2023not,zou2025poisonedrag, liu2024formalizing}, but these attacks treat the LLM as a terminal endpoint rather than a node in a collaborative pipeline. Recent work has begun extending these concerns to multi-agent settings, showing that injected instructions propagate autonomously across agent boundaries~\cite{lee2024prompt,ju2024flooding} and that models resisting direct injection will execute identical payloads when they arrive from a peer agent~\cite{pironti2025dark}. This AI-to-AI trust gap is precisely the implicit trust assumption our boundary framework formalizes. Agents can further engage in covert collusion through channels invisible to standard monitoring~\cite{motwani2024secret}, and multi-step execution amplifies adversarial reach~\cite{debenedetti2024agentdojo,shahroz2025agents}, consistent with the high hop counts we observe under $A_1$ and $A_2$. Defense-oriented frameworks~\cite{zhang2024psysafe,zeng2024autodefense, zhang2024agent} propose mitigations at the model level, yet adaptive attackers continue to bypass them at high success rates~\cite{nasr2025attacker}, reinforcing our finding that model-level intervention is insufficient when the vulnerability is architectural. Broader surveys identify inter-agent trust and traceability as open problems~\cite{deng2025security,he2025emerged, de2025open,raza2026trism,wang2024survey} but do not empirically characterize how these gaps manifest across collaboration boundaries or test whether they are model-specific or structural. Our trace-driven methodology and cross-model evaluation directly address both gaps, and our results confirm that the attack surface is a property of the pipeline architecture rather than the backbone model.

\section{Conclusion}
\label{sec:conclusion} 
In this paper, we formalized boundary verification through three classes governing inter-agent communication, content, identity, and delegation. We hypothesized that adversarial vulnerability in multi-agent LLM pipelines is primarily structural rather than model-dependent. We developed a threat model based on boundary violations and evaluated these vulnerabilities in a controlled multi-agent setting. Our findings show that adversarial behavior arises from how agents interact and exchange information, not from the capability of the underlying models. In current frameworks, message provenance is not enforced, plan-to-execution binding is missing, and identity verification remains weak or absent. As a result, pipelines rely on implicit trust across agents, creating systemic opportunities for adversarial exploitation. Unless these boundaries are explicitly enforced in system design, such vulnerabilities will persist, and improvements in model capability or safety mechanisms alone are not sufficient.


\bibliographystyle{IEEEtran}
\bibliography{IEEEabrv,references}

\end{document}